\documentstyle[12pt,a4]{article}
\def\rbot{r_{\scriptscriptstyle{\bot}}}
\def\half{{{\scriptstyle 1}\over{\scriptstyle 2}}}

\def\botx{{\scriptscriptstyle{\bot}}}

\title{Computer program for the relativistic mean field description of the 
ground state properties of even-even axially deformed nuclei}
\author{P. Ring$^1$, Y.K. Gambhir$^{1,2}$ and G.A. Lalazissis$^{1,3}$\\
\\
$^1$Physik Department, Technische Universit\"at M\"unchen \\
D-85747 Garching, Germany\\
$^2$Physics Department, I.I.T. Powai, Bombay 400076, India\\
$^3$Department of Theoretical Physics, Aristotle University of Thessaloniki,\\
GR 54006 Thessaloniki, Greece.}

\begin{document}
\maketitle
\begin{abstract}
A Fortran program for the calculation of the ground state properties of 
axially deformed even-even nuclei in the relativistic framework is presented. 
In this relativistic mean field (RMF) approach a set of coupled differential
equations namely the Dirac equation with potential terms for the nucleons
and the Glein-Gordon type equations with sources for the meson and the 
electromagnetic fields are to be solved self-consistently. The well tested 
basis expansion method is used for this purpose. Accordingly a set of 
harmonic oscillator basis generated by an axially deformed potential are 
used in the expansion. The solution gives the nucleon spinors, the fields 
and level occupancies, which are used in the calculation of the ground state 
properties.
\end{abstract}

\section {\bf PROGRAM SUMMARY}
\bigskip\bigskip
\noindent
{\it Title of program:} RMFD.f

\bigskip
\noindent
{\it Catalogue number:} ?? 

\bigskip
\noindent
{\it Program available from:} 

\bigskip
\noindent
{\it Licensing provisions:} none

\bigskip
\noindent
{\it Computer for which the program is desinged and others on which it has
been tested:} any Unix work-station or mainframe or PC. The program has been
tested on work stations: DEC, DEC-Alpha , on CDC6600 mainframe 
and on 486 IBM compatible PC's

\bigskip
\noindent
{\it Operating system:} UNIX or VMS or MS-DOS

\bigskip
\noindent
{\it Programming language used:} Fortran 77

\bigskip
\noindent
{\it No. of lines in distributed program, including test data, etc: ca. ??}

\bigskip
\noindent
{\it Keywords:} Relativistic mean field theory, binding energy, nuclear radii,
deformations and densities.

\bigskip
\noindent
{\it Nature of physical problem}

\noindent The Relativistic Mean Field (RMF) theory 
\cite{SW.86,Ser.92}  has been astonishingly
 succesfull \cite{GRT.90,SLR.93,SNR.94,SLH.94,LS.95,LSR.96}
in accurately describing  the nuclear matter properties and the ground 
state properties of finite nuclei spread over the entire periodic table
including those away from the stability line. 
The nucleonic and mesonic degrees of freedom are explicitly included
from the very beginning in the relativistic framework. As a result the
correct spin-orbit splittings emerge automatically. Initially the RMF
theory was applied succesfully to the description of the properties of 
spherical nuclei. As most of the nuclei are deformed, the generalization 
of the solution of the RMF equations for this case is required, 
which is a non trivial task. Therefore a computer program was developed 
\cite{PRB.87,RG.88,GRT.90}  to 
solve the RMF equations, suitable for the calculation of the ground state 
properties of the axially deformed nuclei. The present program is an 
improved version and also is compatible for PC's.  

\bigskip
\noindent
{\it Method of solution}

\noindent In the RMF theory one needs to solve self-consistently a set of
 coupled 
equations namely the Dirac equation with potential terms for the nucleons and
the Klein-Gordon type equations with sources for the mesons and the photon. 
For this purpose we employ the well tested basis expansion method. The bases
used here, are generated by an anisotropic (axially symmetric) harmonic
oscillator potential. The upper and lower components of the nucleon spinors,
the fields as well as the baryon currents and densities are expanded 
separately in these bases. The expansion is truncated so as to include all the 
configurations up to a certain finite value of the  major oscillator shell 
quantum number. In this expansion method the solution of the Dirac equation 
gets reduced to a symmetric matrix diagonalization problem, while that of 
the Klein Gordon equation reduces to a set of inhomogenous equations.      
The solution provides the spinor, fields, and
the nucleon currents and densities (sources of the fields), from which all
the relevant ground state nuclear properties are calculated.   

\bigskip
\noindent
{\it Restrictions on the complexity of the problem}

\noindent The present version is applicable to even-even nuclei due to the
imposition of time reversal inavariance and charge conservation.
The program can be modified  for the general case including that of odd mass 
nuclei by incorporating the additional currents arising due to time 
reversal breaking.
 
\bigskip
\noindent
{\it Typical running time:}

\noindent From 30 minutes to several hours depending upon the computer for
the general case. However, the computer time will considerably increase if 
one wishes to include higher number of shells. 

\noindent  

\section{\bf LONG WRITE-UP}

\subsection{\bf RMF equations}

The basic Ansatz of the RMF theory is a Lagrangian density
\cite{SW.86,Ser.92} where nucleons are described as Dirac
particles which interact via the exchange of various
mesons. The Lagrangian density considered is written in the form:
\begin{equation}
\begin{array}{rl}
{\cal L} &=
\bar \psi (i\rlap{/}\partial -M) \psi +
\,{1\over2}\partial_\mu\sigma\partial^\mu\sigma-U(\sigma)
-{1\over4}\Omega_{\mu\nu}\Omega^{\mu\nu}+\\
\                                        \\
\ & {1\over2}m_\omega^2\omega_\mu\omega^\mu
-{1\over4}{\vec R}_{\mu\nu}{\vec R}^{\mu\nu} +
 {1\over2}m_{\rho}^{2} \vec\rho_\mu\vec\rho^\mu
-{1\over4}F_{\mu\nu}F^{\mu\nu} \\
\                              \\
 &  g_{\sigma}\bar\psi \sigma \psi~
     -~g_{\omega}\bar\psi \rlap{/}\omega \psi~
     -~g_{\rho}  \bar\psi 
      \rlap{/}\vec\rho
      \vec\tau \psi
     -~e \bar\psi \rlap{/}A \psi
\end{array}
\end{equation}
The meson fields included are the isoscalar $\sigma$ meson,
the isoscalar-vector $\omega$ meson and the
isovector-vector $\rho$ meson. The latter provides the
necessary isospin asymmetry. 

 The arrows in Eq.
(1) denote the isovector quantities.  The Lagrangian
contains also a non-linear scalar self-interaction of the
$\sigma$ meson.
\begin{equation}
U(\sigma)~={1\over2}m_{\sigma}^{2} \sigma^{2}~+~
{1\over3}g_{2}\sigma^{3}~+~{1\over4}g_{3}\sigma^{4}
\end{equation} 
This term is important for appropriate
description of surface properties \cite{BB.77}.
M, m$_{\sigma}$, m$_{\omega}$ and m$_{\rho}$ are the nucleon-,
the $\sigma$-, the $\omega$- and the $\rho$-meson masses
respectively, while g$_{\sigma}$, g$_{\omega}$, g$_{\rho}$
and e$^2$/4$\pi$ = 1/137 are the corresponding coupling
constants for the mesons and the photon.
The field tensors of the vector mesons and of the
electromagnetic fields take the following form:
\begin{eqnarray}
\Omega^{\mu\nu} =& \partial^{\mu}\omega^{\nu}-\partial^{\nu}\omega^{\mu}\\
\vec R^{\mu\nu} =& \partial^{\mu}\vec\rho^\nu-\partial^{\nu}\vec\rho^\mu\\
F^{\mu\nu} =& \partial^{\mu}A^{\nu}-\partial^{\nu}A^{\mu}
\end{eqnarray}

The variational principle gives the equations of motion.
The mean field approximation is introduced at this stage by
treating the fields as the
c-number or classical fields. This  results into a set of coupled equations 
namely the Dirac equation with potential terms for the nucleons and
the Klein-Gordon type equations with sources for the mesons and the photon.
For the static case, along with  the time reversal invariance and
charge conservation the equations get simplified. The resulting equations,
known as RMF equations have the following form. 

 The Dirac equation for the nucleon: 
\begin{equation}
\{ -i{\mbox \boldmath \alpha}{\mbox \boldmath \nabla} + 
V({\bf r}) + \beta [ M +S({\bf r}) ] \} 
\psi_i~=~\varepsilon_i\psi_i,
\end{equation}
where $V({\bf r})$ represents the $vector$ potential:
\begin{equation}
V({\bf r}) = g_{\omega} \omega_{0}({\bf r}) + g_{\rho}\tau_{3} {\bf {\rho}}
_{0}({\bf r}) + e{1+\tau_{3} \over 2} A_{0}({\bf r}),
\end{equation}
and $S({\bf r})$ is the $scalar$ potential:
\begin{equation}
S({\bf r}) = g_{\sigma} \sigma({\bf r}) 
\end{equation}
the latter contributes to the effective mass as:
\begin{equation}
M^{\ast}({\bf r}) = M + S({\bf r}).
\end{equation}

 The Klein-Gordon equations for the meson and the electromagnetic fields 
with the nucleon densities as sources:
\begin{eqnarray}
\{ -\Delta + m_{\sigma}^{2} \}\sigma({\bf r})
&=& -g_{\sigma}\rho_{s}({\bf r})
-g_{2}\sigma^{2}({\bf r})-g_{3}\sigma^{3}({\bf r})\\
\{ -\Delta + m_{\omega}^{2} \} \omega_{0}({\bf r})
&=& g_{\omega}\rho_{v}({\bf r})\\
\{ -\Delta + m_{\rho}^{2} \}\rho_{0}({\bf r})
&=& g_{\rho} \rho_{3}({\bf r})\\
 -\Delta A_{0}({\bf r}) &=& e\rho_{c}({\bf r})
\end{eqnarray}

The corresponding densities are:
\begin{equation}
\begin{array}{ll}
\rho_{s} =& \sum\limits_{i=1}^{A} n_{i} \bar\psi_{i}~\psi_{i}.\\
\             \\
\rho_{v} =& \sum\limits_{i=1}^{A} n_{i} \psi^{+}_{i}~\psi_{i}.\\
\             \\
\rho_{3} =& \sum\limits_{p=1}^{Z} n_{i} \psi^{+}_{p}~\psi_{p}~-~
\sum\limits_{n=1}^{N} n_{i} \psi^{+}_{n}~\psi_{n}.\\
\                    \\
\ \rho_{c} =& \sum\limits_{p=1}^{Z} n_{i} \psi^{+}_{p}~\psi_{p}.
\end{array}
\end{equation}
Here the sums are taken over the particle states only.
This implies that the  contributions from negative-energy
states are neglected ($no$-$sea$ approximation), i.e. the
vacuum is not polarized.
The $\pi$ meson does not contribute in the present relativistic mean field 
(Hartree) approxiamtion because of its pseudo nature. The occupation
number $n_i$ is introduced to account for pairing which is important for
open shell nuclei. In the absence of pairing  it takes the value one (zero) 
for the levels below (above) the Fermi surface. In the presence of pairing 
the partial occupancies ($n_i$) are obtained in the constant gap approximation
(BCS) through the well known expression:
\begin{equation} 
n_i =\ {1\over 2}(\, 1\  - \ 
{{\varepsilon_i - \lambda}\over
\sqrt{(\varepsilon_i - \lambda)^2+\Delta^2}})
\end{equation} 
The 
$\varepsilon_{i}$ is the single-paricle energy for the state $i$  and 
chemical potential or Fermi energy
$\lambda$ for protons (neutrons) is obtained from the requirement
\begin{equation}
\begin{array}{rl}
\displaystyle \sum_{i} n_i  = & \mbox{the number of protons (Z)} \\
\               & (\mbox{the number of neutrons (N)})
\end{array}
\end{equation}
The sum is taken over protons (neutrons) states.
The gap parameter $\Delta$ is calculated from the observed odd-even mass
differences. In the absence of experimental masses it can inferred from the
extrapolation of the masses given by any of the avalaible mass  formulae 
(e.g from ref. \cite{MNM.95})

The above set of equations (6,10-13) are to be solved self-consistently. 
For this purpose one starts with an initial guess of the fields 
(e.g. generated by axially deformed Woods-Saxon potential) to calculate 
the potential terms (Eqs. (7,8)) appearing in the Dirac equation (Eq. (6)).
The Dirac equation is solved with these potentials 
terms to yield the nucleon spinors which in term are used to obtain the
sources (densities). The meson and photon equations are 
then solved with these sources
to get a new set of fields to be used for the calculation of new
potential terms. The Dirac equation is then solved with the new potentials to
get the spinors again to be used to obtain the new sources 
for the meson fields. This iterative procedure is continued till 
the converegence upto the desired accuracy is achieved.   

\subsection{\bf Axially symmetric case} 
For the axially symmetric deformed shape 
the rotational symmetry is broken and therefore, the total angular momentum 
$j$ is no longer a  good quantum number. However, the densities are 
still invariant with respect to a rotation around the symmetry axis, which 
is taken to be the z-axis. It then turns out to be useful to work with
the cylindrical coordinates

\begin{equation}
x = \rbot \cos\varphi ,\quad y = \rbot \sin\varphi\quad 
{\rm and}\quad z.
\end{equation}

For such nuclei the Dirac equation can be reduced to a coupled set of
partial differential equations in the two variables $z$ and $\rbot$. In
particular, the spinor $\psi_i$ with the index $i$ is now characterized
by the quantum numbers
\begin{equation}
\Omega_i, \pi_i, t_i
\end{equation}
where $\Omega_i$ is the eigenvalue of the symmetry
operator $j_{z_{i}}$ (the projection of $j_{i}$ on the z-axis), $\pi_i$ is 
the parity and $t_i$ is the isospin. The spinor can be written in the form:
\smallskip
\begin{equation}
\psi_i ({\bf r},t)\ =\ 
\pmatrix{f_i({\bf r})\cr ig_i({\bf r})\cr}\ =\ 
{1\over\sqrt{2\pi}}
\pmatrix{
 f_i^+(z,\rbot)\ e^{i(\Omega_i-1/2)\varphi}\cr
 f_i^-(z,\rbot)\ e^{i(\Omega_i+1/2)\varphi}\cr
ig_i^+(z,\rbot)\ e^{i(\Omega_i-1/2)\varphi}\cr
ig_i^-(z,\rbot)\ e^{i(\Omega_i+1/2)\varphi}\cr}\ \chi_{t_i}(t)
\end{equation}

\medskip\noindent
The four components $f_i^\pm(\rbot ,z)$ and $g_i^\pm(\rbot ,z)$ obey
the Dirac equations
\begin{eqnarray}
(M^* + V) f_i^+\ +\ \partial_z g_i^+\  
+\ (\partial_{\rbot} + {{\Omega+{1\over 2}}\over {\rbot}}) g_i^-\ 
&&=\ \epsilon_i\ f_i^+
\\ 
(M^* + V) f_i^-\ -\ \partial_z g_i^-\  
+\ (\partial_{\rbot} - {{\Omega-{1\over 2}}\over {\rbot}}) g_i^+\ 
&&=\ \epsilon_i\ f_i^-
\\
(M^* - V) g_i^+\ +\ \partial_z f_i^+\  
+\ (\partial_{\rbot} + {{\Omega+{1\over 2}}\over {\rbot}}) f_i^-\ 
&&=\ -\epsilon_i\ g_i^+
\\ 
(M^* - V) g_i^-\ -\ \partial_z f_i^-\  
+\ (\partial_{\rbot} - {{\Omega-{1\over 2}}\over {\rbot}}) f_i^+\ 
&&=\ -\epsilon_i\ g_i^-
\end{eqnarray}
\medskip\noindent
For each solution with positive $\Omega$: 
\begin{equation}
\psi_i\ \equiv \{ f_i^+,f_i^-,g_i^+,g_i^-,\Omega_i\}
\end{equation}
we have the time reversed solution with the same energy
\begin{equation}
\psi_{\bar\iota}\ =\ {\it T}\psi_i\  
\equiv\ \{ -f_i^-,f_i^+,g_i^-,-g_i^+,-\Omega_i\}
\end{equation}
with the time reversal operator ${\it T} = i\sigma_y {\it K}$ ({\it K}
being the complex conjugation). For nuclei with time reversal symmetry,
the contributions to the densities of the two time reversed states $i$
and ${\bar i}$ are identical. Therefore, we find the densities
\begin{equation}
\rho_{s,v}\ =\ 2 \sum_{i>0} n_i 
\bigl( 
(\mid f_i^+\mid^2 + \mid f_i^-\mid^2)\ 
\mp\ (\mid g_i^+\mid^2 + \mid g_i^-\mid^2 )
\bigr)
\end{equation}
and, in a similar way, $\rho_3$ and $\rho_c$.
The sum $i>0$ runs only over the states with positive $\Omega_i-
$values. These densities serve as sources for the fields $\phi$ =
$\sigma$, $\omega^0$ $\rho^0$ and $A^0$, which are determined by the
Klein-Gordon equation in cylindrical coordinates:
\begin{equation}
\bigl( -{1\over \rbot}\partial_{\rbot} \rbot \partial_{\rbot}
\ -\ {\partial_z}^2\ +\ {m_\phi}^2 \bigr) \phi (z,\rbot)
\ =\ s_\phi(z,\rbot)
\end{equation}
The inhomogeneous parts $s_\phi$ are given by
\begin{equation}
s_\phi (z,\rbot) = \ \cases{  
- g_\sigma \rho_s(z,\rbot) - g_2 \sigma^2(z,\rbot) - g_3 \sigma^3(z,\rbot) 
                    &\qquad for the $\sigma-$field\cr 
\ g_\omega \rho_v(z,\rbot) &\qquad for the $\omega-$field\cr
\ g_\rho \rho_3(z,\rbot)  &\qquad for the $\rho-$field\cr
\ e\rho_p(z,\rbot)&\qquad for the Coulomb field\cr}
\end{equation}

\subsection{Solution of the RMF equations}

For the solution of thr RMF equations we use the basis expansion method.
We follow closely the details, presentation and the notation of
ref. \cite{GRT.90}.
For the axially symmetric case we expand the spinors 
$f_i^{\pm}$ and $g_i^{\pm}$ in Eqs. (20-23)
in terms of the eigenfunctions of a deformed axially symmetric
oscillator potential 
\begin{equation}
V_{osc}(z,\rbot)\quad =\quad 
\half M \omega_z^2 z^2\ +\ 
\half M \omega_{\botx}^2 \rbot^2  
\end{equation}
Imposing volume conservation, the two oscillator frequencies 
$\hbar\omega_{\botx}$ and $\hbar\omega_z$ can be expressed in terms of
a deformation parameter $\beta_0$:
\begin{eqnarray}
\hbar\omega_z\ &&=\ \hbar\omega_0\ 
\exp(-{\scriptstyle \sqrt{5\over{4\pi}}}\beta_0) 
\\
\hbar\omega_{\botx}\ &&=\ \hbar\omega_0\ 
\exp(+{\scriptstyle {1\over 2}\sqrt{5\over{4\pi}}}\beta_0) 
\end{eqnarray}

The corresponding oscillator length parameters are
\begin{equation}
b_z\ =\ \sqrt{\hbar\over{M\omega_z}} 
\quad {\rm and} \quad 
b_{\botx}\ =\ \sqrt{\hbar\over{M\omega_{\botx }}}
\end{equation}
Because of volume conservation, we have $b_{\botx }^2 b_z = b_0^3$. The
basis is now determined by the two constants $\hbar\omega_0$ and
$\beta_0$. The
eigenfunctions of the deformed harmonic oscillator potential are
characterized by the quantum numbers 
\begin{equation}
\mid\alpha >\ =\ \mid n_z,n_r,m_l,m_s>
\end{equation}
where $m_l$ and $m_s$ are the components of the orbital angular
momentum and of the spin along the symmetry axis. The eigenvalue of
$j_z$, which is a conserved quantity in these calculations, is 
\begin{equation}
\Omega\ =\ m_l + m_s.
\end{equation}
The parity is given by
\begin{equation}
\pi\ =\ (-)^{n_z+m_l}
\end{equation}
The eigenfunctions of the deformed harmonic oscillator can be
written explicitly as
\begin{eqnarray}
\Phi_\alpha (z,\rbot,\varphi,s,t)\ = && \ 
\phi_{n_z}(z)\ \phi_{n_r}^{m_l}(\rbot )  
{ {\scriptstyle 1}\over{\scriptstyle{\sqrt{2\pi}}}} e^{im_l\varphi}
\chi_{ms}(s)\chi_{t_{\alpha}}(t)\\
\ &&=  \Phi_\alpha ({\bf r},s)\, \chi_{t_{\alpha}}(t)
\end{eqnarray}
with
\begin{eqnarray}
\phi_{n_z}(z)\ \ &&=\ {{N_{n_z}}\over{\sqrt{b_z}}}\ H_{n_z}(\zeta)
\, e^{-\zeta^2 /2}
\\
\phi_{n_r}^{m_l}(\rbot )\ &&=\ { {N_{n_r}^{m_l} } \over {b_{\botx}}}\ 
\sqrt{2}\, \eta^{m_l/2}\, L_{n_r}^{m_l}(\eta )\, e^{-\eta /2}
\end{eqnarray}
and
\begin{equation}
\zeta\ =\ z/b_z,\qquad \eta\ =\ \rbot^2/b_\botx^2
\end{equation}
The polynomials $H_n(\zeta)$ and $L_n^m(\eta)$ are Hermite polynomials 
and associated Laguerre polynomials as defined in ref. \cite{AS.70}
The normalization constants are given by
\begin{equation}
N_{n_z}\ =\ {1\over\sqrt{\sqrt{\pi}2^{n_z}n_z!}}
\quad {\rm and} \quad
N_{n_r}^{m_l}\ =\ \sqrt{{n_r!}\over{(n_r+m_l)!}} 
\end{equation}
\medskip
In order to evaluate the matrix elements, we also need the polynomials 
${\tilde L}_{n_r}^{m_l}$ and ${\tilde H}_{n_z}$ defined by the
derivatives
\begin{eqnarray}
\partial_z \phi_{n_z}(z)\ 
&&=\ {{N_{n_z}}\over{b_z^{3/2}}}\, {\tilde H}_{n_z}(\zeta)\, e^{-\zeta^2 /2}
\\
\partial_{\rbot } \phi_{n_r}^{m_l}(\rbot )\ 
&&=\ {{N_{n_r}^{m_l}}\over{b_{\botx }^2}}\, \sqrt{2} \eta^{(m_l-1)/2}
\, {\tilde L}_{n_r}^{m_l} (\eta)\, e^{-\eta /2}
\end{eqnarray}
which can be calculated from the recursion relations
\begin{eqnarray}
{\tilde H}_{n_z}(\zeta)\ &&=\ \zeta H_{n_z}(\zeta) - H_{n_z+1}(\zeta) 
\\
{\tilde L}_{n_r}^{m_l}(\eta)\ &&=\ (2n_r+m_l-\eta)\, L_{n_r}^{m_l}(\eta)
-2(n_r+m_l)\, L_{n_r-1}^{m_l} (\eta) 
\end{eqnarray}
The solutions of the Dirac equation in the axially symmetric case have
only the good quantum numbers $\Omega$ and $\pi$ and we use the
expansion
\begin{eqnarray}
f_i({\bf r},s,t)\ &&=\ {1\over\sqrt{2\pi}}
\pmatrix{
f_i^+(z,\rbot)\ e^{i(\Omega - 1/2)\varphi} \cr
f_i^-(z,\rbot)\ e^{i(\Omega + 1/2)\varphi} \cr}
\ =\ \sum_\alpha^{\alpha_{max}} f_\alpha^{(i)} 
\, \Phi_\alpha ({\bf r},s)\, \chi_{t_i}(t)
\\
g_i({\bf r},s,t)\ &&=\ {1\over\sqrt{2\pi}}
\pmatrix{
g_i^+(z,\rbot)\ e^{i(\Omega - 1/2)\varphi} \cr
g_i^-(z,\rbot)\ e^{i(\Omega + 1/2)\varphi} \cr}
\ =\ \sum_{\tilde\alpha}^{{\tilde\alpha}_{max}} g_{\tilde\alpha}^{(i)} 
\, \Phi_{\tilde\alpha} ({\bf r},s)\, \chi_{t_i}(t)
\end{eqnarray}
To avoid the appearence of the spurious states \cite{GRT.90,GR.93} 
the quantum numbers 
$\alpha_{max}$ and ${\tilde\alpha}_{max}$ are chosen in such a way that
the corresponding major quantum numbers $N=n_z+2n_\rho+m_l$ are not
larger than $N_F+1$ for the expansion of the small components, and not
larger than $N_F$ for the expansion of the large components. 
\medskip
The Dirac equation reduces to a symmetric matrix diagonalization problem:
\smallskip
\begin{equation}
\pmatrix{
 {\cal A}_{\alpha,\alpha '}
&{\cal B}_{\alpha,{\tilde\alpha}'}\cr
 {\cal B}_{{\tilde\alpha},\alpha '}
&-{\cal C}_{{\tilde\alpha},{\tilde\alpha}'}\cr}
\pmatrix{
f_{\alpha '}^{(i)}\cr 
g_{{\tilde\alpha}'}^{(i)}\cr}
\quad =\quad \epsilon_i 
\pmatrix{
f_{\alpha}^{(i)}\cr 
g_{\tilde\alpha}^{(i)}\cr}
\end{equation}
\smallskip\noindent
of the dimension $\alpha_{max}+{\tilde\alpha}_{max}$. The matrix
elements 
${\cal A}_{\alpha\alpha '}$, 
${\cal B}_{\alpha\alpha '}$ and
${\cal C}_{\alpha\alpha '}$ are given by:
\begin{eqnarray}
\pmatrix{
{\cal A}_{\alpha\alpha '}\cr
{\cal C}_{\alpha\alpha '}\cr}&&
\ =\ \delta_{m_l {m_l}'} \delta_{m_s m_s'}\quad
N_{n_r}^{m_l} N_{n_z} N_{{n_r}'}^{{m_l}'} N_{{n_z}'}  
\int\limits_0^{\infty}d\eta e^{-\eta} \eta^{m_l}
L_{n_r}^{m_l}(\eta ) L_{{n_r}'}^{m_l}(\eta )
\\
&&\times\int\limits_0^{\infty}d\zeta e^{-\zeta^2}
H_{n_z}(\zeta ) H_{{n_z}'}(\zeta )
\ \left( 
    M^*(b_z\zeta,b_{\botx }\sqrt{\eta})\ 
\pm\  V(b_z\zeta,b_{\botx }\sqrt{\eta})
\right) 
\end{eqnarray}
\begin{eqnarray}
{\cal B}_{\alpha\alpha'}\ 
&&=\ \delta_{m_l {m_l}'} \delta_{m_s m_s'}
\delta_{n_r {n_r}'} 
{ {(-)^{\half - m_s}} \over {b_z} }
\left(\delta_{{n_z}'n_z +1}\sqrt{{{n_z}'}\over2}
              -\delta_{n_z {n_z}'+1}\sqrt{{n_z}\over2}\right)
\\
&&+\delta_{m_l {m_l}'}\delta_{n_z {n_z}'}
{ {N_{n_r}^{m_l} N_{{n_r}'}^{{m_l}'} }\over{b_{\botx }}}
\Biggl\{
\\
&&\delta_{m_s' m_s+1}
\int\limits_0^{\infty}d\eta e^{-\eta} \eta^{m_l-1/2}
L_{n_r}^{m_l}(\eta ) 
\left( {\tilde L}_{{n_r}'}^{m_l}(\eta) - m_l L_{{n_r}'}^{m_l}(\eta)
\right)
\\
+\ &&\delta_{m_s m_s'+1} 
\int\limits_0^{\infty}d\eta e^{-\eta} \eta^{m_l-1/2}
L_{n_r}^{m_l}(\eta ) 
\left({\tilde L}_{{n_r}'}^{m_l}(\eta) + (m_l+1)L_{{n_r}'}^{m_l}(\eta)
\right)
\Biggr\}
\end{eqnarray}
\medskip\noindent
In the next step, we calculate the density matrices in the shell model
space:
\begin{equation}
\rho_{\alpha\alpha'}^{s,v}\ =\ 
2 \sum_{i>0} n_i (f_\alpha^{(i)} f_{\alpha '}^{(i)}\ \mp\ 
                  g_\alpha^{(i)} g_{\alpha '}^{(i)})
\end{equation}
and in coordinate space:
\begin{eqnarray}
\rho_{s,v}(z,\rbot)
\ &&=\ {1\over{b_0^3}}{1\over\pi} e^{-\zeta^2-\eta}\ 
\sum_{\alpha\alpha '}\quad\rho_{\alpha\alpha '}^{s,v} 
\quad\delta_{m_l {m_l}'}\\
&&\quad\times\ N_{n_z} N_{{n_z}'} 
H_{n_z}(\zeta) H_{{n_z}'}(\zeta)\  
N_{n_r}^{m_l} N_{{n_r}'}^{m_l}
\, \eta^{m_l}
\, L_{n_r}^{m_l}(\eta) L_{{n_r}'}^{m_l}(\eta).
\end{eqnarray}
Similarly, one obtains $\rho_3$ and $\rho_c$. These are the sources for
the solution of the Klein-Gordon equations. The
fields of massive mesons are also solved by the expansion in a deformed
oscillator basis. Here for computational and numerical convenience we use 
the same deformation parameter $\beta_0$ as in
Eqs (32,33), and take the oscillator length 
$b_B = b_0/\sqrt{2}$ :
\begin{equation}
\phi (z,\rbot)\ =\ {1\over{b_B^{3/2}}} e^{-\zeta^2/2-\eta/2}
\sum_{n_z n_r}^{N_B} \phi_{n_z n_r} 
N_{n_z}\, H_{n_z} (\zeta)\ \sqrt{2}\, L_{n_r}^0 (\eta)
\end{equation}
with $\zeta = \sqrt{2} z/b_z$ and $\eta = 2 \rbot ^2 /b_{\botx }^2$.
Inserting this ansatz into the Klein-Gordon equation, we obtain an
inhomogeneous set of linear equations
\begin{equation}
\sum_{{n_z}'{n_r}'}^{N_B} {\cal H}_{n_z n_r {n_z}' {n_r}'}\
\phi_{{n_z}' {n_r}'}
\ =\ s_{n_z n_r}^{\phi}
\end{equation}
with the matrix elements
\begin{eqnarray}
{\cal H}_{n_z n_r {n_z}' {n_r}'}\ =\ 
&&\bigl( 
-{2\over b_z^2} (n_z + 1/2)\ -\ {2\over b_\botx^2} (2n_r +1)
\ +\ m_\phi^2 \bigr) \delta_{n_r {n_r}'} \delta_{n_z {n_z}'}
\\
+\ &&{2\over b_z^2} 
\bigl( \half\sqrt{(n_z +1){n_z}'}\ \delta_{{n_z}'n_z +1}
\ +\   \half\sqrt{({n_z}'+1)n_z }\ \delta_{n_z {n_z}'+1} \bigr)
\\
-\ &&{2\over b_\botx^2} 
\bigl( {n_r}'\ \delta_{{n_r}' n_r+1}\ +\ n_r\ \delta_{n_r {n_r}'+1}
\bigr)
\end{eqnarray}
For the Coulomb field, due to its long range character we cannot use the
oscillator basis expansion method. Therefore for this case the standard
Green's function method is used as is discussed in the appendix 
of ref. \cite{VAUT.73} 

In the present case with time reversal symmetry and pairing the total binding
energy is given by the sum of various individual contributions:
\begin{equation}
E =\ 
E_{part}+E_\sigma + E_\omega + E_\rho + E_{c} + E_{pair}
+ E_{CM} - A M
\end{equation}
with
\begin{eqnarray}
E_{part}\ &&=\ \sum_i {n_i}^2 \epsilon_i\\
E_{\sigma L}\ &&=\ -{{g_\sigma}\over2}\int d^3 r\,
\rho_s({\bf r})\sigma({\bf r})\\
E_{\sigma NL}\ &&=\ -{1\over2} \int d^3 r\,
\bigl\{ {2\over3} g_2\sigma({\bf r})^3\ +\ 
        {1\over2} g_3\sigma({\bf r})^4 \bigr\}\\
E_{\omega}\ &&=\ -{{g_\omega}\over2}\int d^3 r\,
\rho_v({\bf r})\omega^0({\bf r})\\
E_{\rho}\ &&=\ -{{g_\rho}\over2}\int d^3 r\,
\rho_3({\bf r})\rho^{00}({\bf r})\\
E_c\ &&=\ -{{e^2}\over{8\pi}} \int d^3 r\,
\rho_c({\bf r})A^0({\bf r})\\
E_{pair}\ &&=\ - \Delta \sum_{i} \sqrt{n_{i}(1-n_{i})}\\
E_{CM}\ &&=\ -{3\over 4} \hbar\omega_0
\ =\ -{3\over 4} 41 A^{-{1\over 3}}
\end{eqnarray}
In the calculation of the pairing energy we use a {\it pairing window}, 
i.e. the sum over $i$ in Eq. (69) is only
extended up to the level where
$\epsilon_i - \lambda \leq 2 (41 A^{-{1\over3}})$. The factor $2$ has
been determined so as to reproduce the pairing correlation energy (Eq. (69))
for neutrons in the nucleus $^{118}$Sn calculated by using the Gogny force
\cite{DG.80}.

\medskip
The charge radius is calculated using the following formula:
\begin{equation}
r_c\ =\ \sqrt{r_p^2 + 0.64} \qquad(fm) 
\end{equation}
\medskip
The factor 0.64 in Eq. (71) accounts for the finite size effects of the proton.

The quadrupole $Q_{n,p}$ and hexadecupole $H_{n,p}$ moments 
for neutrons and protons are calculated using the expressions: 
\begin{equation}
Q_{n,p}\ =\ < 2 r^2 P_2(\cos\theta) >_{n,p}\ =\ 
<2z^2 - x^2 - y^2>_{n,p}
\end{equation}
and
\begin{equation}
H_{n,p}\ =\ <  r^4 Y_{40}(\theta) >_{n,p}\ =\ 
\sqrt{9\over{64\pi}} <8z^4-24z^2(x^2+y^2)+3(x^2+y^2)^2>_{n,p}
\end{equation}

The conventional deformation parameter $\beta$ is obtained from the
calculated qua\-dru\-pole moments through
\begin{equation}
Q = Q_n + Q_p = \sqrt{{16\pi}\over 5} {3\over{4\pi}} A R_0^2 \beta. 
\end{equation}
with $R_0 = 1.2 A^{1\over3}~(fm)$. 

\subsection{\bf Calculation of ground state properties }

The present program has been used to calculate the
ground propetries of axially deformed even-even nuclei spread over the
entire periodic table encompasing also the nuclei far away from the 
stability line.

For carrying  out explicit numerical
calculations one requires the following input information:

 i) The baryon and the meson masses and the coupling constants of the 
meson fields to the nucleons appearing in the Lagrangian (Eq.(1))

 ii) The number of oscillator shells $N_F$ and $N_B$, i.e.
the cutoff major oscillator shell quantum number up to which the Dirac 
spinors (Fermion wavefunctions)
and meson fields (describing the Bosonic degrees of freedom), and so also
the densities, are expanded.

 iii) The basis parameters $\hbar\omega_0$ and $\beta_0$ used
for the expansion of the Dirac spinors and the fields. Following 
ref. \cite{GRT.90} we fix  for fermions $\hbar\omega_0 = 41 A^{-1/3}$ 
and $\beta_0$ can be taken to be any reasonable value 
(preferably close to the experimental value).

As stated before the oscillator size parameter for
the meson fields (and the densities) are fixed at $1/\sqrt{2}$ times
the corresponding Fermion parameters. The deformation parameter of the
oscillator basis $\beta_0$ has been chosen to be identical for the Dirac
spinors and the meson fields. This simplifies the calculations and
avoids the need for additional parameters. 

\medskip
For illustration we present here the results of calculations for the
ground state properties of Sr nuclei over a wide range of isospin 
\cite{LS.95}, namely 
Sr isotopes with mass numbers A=70 up to A=110.  
The calculations have been carried out using the Lagrangian parameter set
NL-SH, which provides good results on both sides of the stability 
line \cite{SNR.93}. The NL-SH set of parameters are listed in Table 1.

The number of shells taken into account in the expansion is 12 for 
both fermions and bosons ($N_{F}$=$N_{B}$=12). 
It should be noted that for convergence reasons 14 shells were also considered.
It was observed that, there is hardly any difference between these two sets of
results. This observation also holds for nuclei in the entired periodic table.
Therefore, it is sufficient to consider 12 shells in practical calculations. 
However, for superheavy nuclei probably one has to include higher number of
shells.  

For open shell nuclei, pairing has been included using the BCS formalism. 
In the BCS calculations we have used constant pairing gaps,
which are taken from the empirical particle separation energies \cite{audi93} 
of neighbouring nuclei.

\medskip
Now we present and discuss some of the calculated results.
Fig. 1 shows the binding energy per nucleon (E/A) for 
Sr isotopes.  
The empirical values taken from the 1993 Atomic Mass Evaluation 
Tables \cite{audi93} (expt.) are also shown. The figure 
also includes the predictions of the recent finite-range droplet 
model \cite{MNM.95} (FRDM) and of the extended Thomas-Fermi with Strutinsky 
Integral (ETF-SI) model \cite{APD.95} for comparison. The parabolic 
shapes of the binding energy per nucleon emerges nicely.
The minimum in the binding energy is observed at the magic neutron number 
N=50 in the RMF as well as in the mass models.The calculated  RMF binding 
energies agree closely to the predictions of the FRDM and ETFSI models 
within 1-2 MeV. The RMF theory predicts binding energies which are in 
accord with the empirical values in almost all the cases, with deviations
$\leq$ 0.5\%. It should be noted that similar type of agreement has also 
been obtained in other mass regions. 

We have performed calculations in the RMF theory for both the prolate 
and oblate configurations. The deformations 
of nuclei have been obtained from the relativistic Hartree 
minimization. We show in Fig. 2 the quadrupole deformation $\beta_2$
for the shape corresponding to the lowest energy. The predictions of 
FRDM and ETF-SI are also shown for comparison. It is seen that the RMF theory 
gives a well-defined prolate shape for lighter 
isotopes. Further, an addition of a few neutrons below the closed neutron
shell leads to an oblate shape. This shape
turns into spherical ones as nuclei approach the magic neutron number
N=50. Nuclei above this magic number revert again to the prolate shape
in the RMF theory. Thus, a shape transition
from prolate-oblate-spherical-prolate is followed.

In addition to the lowest minimum, several isotopes exhibit a second 
minimum, thus implying a shape-coexistence, i.e., the prolate and the 
oblate shapes differ in the energy only by a few hundred keV. These nuclei 
have been shown by squares surrounding the black circles.
The calculations predict two minima for several heavy Sr nuclei,
the prolate shape results being a few hundreds keV lower in 
energy than the oblate one. This is displayed in Fig. 3, where the 
difference in the ground-state binding energy of the oblate and the 
prolate configurations is shown. It can be noticed that Sr isotopes 
beginning with A=92 acquire a prolate shape predominantly. For nuclei 
close to A=98, the prolate shape is lower than the oblate one only 
by about 300 keV. With a further increase in the neutron number the 
Sr isotopes take up the prolate minimum, the oblate shape being 
about 600-800 keV higher. The undulation in the prolate-oblate energy 
differences of neutron-rich Sr isotopes is a noteworthy feature of 
the RMF prediction. 

The $\beta_2$ values from NL-SH, FRDM and ETF-SI is shown in Table 2.
The $\beta_2$ values for the second minimum obtained for Sr isotopes
with the force NL-SH are shown in the parentheses. A comparison  shows that 
the three approaches provide values close to each other.  
The experimental quadrupole deformations obtained from BE(2) values taken 
from \cite{raman87} are shown in the last columns of the table. It may be 
noted that these empirical $\beta_2$ values do not indicate any signature 
as to ascertain the shape of a given nucleus. The absolute values, however, 
do compare well with the RMF predictions. 

In Fig.4 the rms charge and neutron radii of Sr nuclei are presented. 
It is seen that on going from the lighter isotopes to the heavier ones
the charge radii exhibit a decreasing trend upto the magic isotope, that is
the lighter isotopes have higher charge radii than the heavier closed neutron-
shell nucleus. The charge radii for nuclei heavier than the 
closed neutron-shell start increasing with addition of neutrons. The 
neutron radii, on the other hand, also show a kink about the neutron shell 
closure. However, the neutron radius for lighter isotopes in these chains 
is not higher than that of the closed-shell nucleus.  

In Fig. 5 we show   the isotope shifts $r^{2}_{c}(A) - r^{2}_{c}(ref))$
for Sr nuclei calculated with respect to a reference nucleus ($^{88}$Sr).
The empirical data obtained from atomic laser spectroscopy \cite{ott88,buch90}
are also shown. The experimental data for Sr nuclei exhibit a kink about 
the magic neutron number. This kink about the closed-shell is a characteristic
feature of isotope shifts in many nuclei \cite{ott88}. 
A solution to this problem has eluded since long. It can be seen from the
figure that the RMF theory is successful in reproducing this kink. 
We would like to stress that the RMF theory  provides a first-ever microscopic
description of anomalous isotopic shifts in Sr nuclei.
Such an anomalous behaviour is a generic feature of deformed nuclei 
which include almost all isotopic chains in the rare-earth region \cite{ott88}.
It is to be remarked that here also the RMF theory is  shown to have a 
remarkable success \cite{LSR.96}. This is amply demonstrated through Fig.6 
where the isotope-shift of rare-earth nuclei with atomic numbers
ranging from Z=60 upto Z=70 are shown.

In short the RMF theory is highly successful in describing the ground state 
properties ( e.g binding energies, nuclear radii, the deformations, isotope
shifts etc). The agreement with the experiment is really remarkable.
It is to be stressed that the RMF theory uses limited number
of parameters as compared to other theories and models and at the same
times it provides a consistent and unified 
description of the ground state properties of nuclei over range of isospin.

\section{ \bf Program structure and test run}

The code consists of the fortran program and two additional files: DIZ.PAR and
DIZ.DAT. The file DIZ.PAR contains the relevant information regarding 
the dimensions, depending upon the number of oscillator shells to be included. 
The standard form of DIZ.PAR is for 12 shells both for fermions and bosons 
($N_{F}$=$N_{B}$=12).  
The input file DIZ.DAT  provides the necessary information 
relevant to the specific case being calculated (the details are attached). 
The program runs interactively. After three iterations the user has to provide
as input the number of iterations and the value of the parameter xmix.
As an example for hundred iterations and the value of xmix=0.3 the user should
write -100 and in the next 0.3.    
The main program DIZ calls various subroutines reading the data and performing
the execution. The operation consists essentially of two parts. 
The first part using DIZ.DAT starts the
program, initializes and  generates all the relevant information. 
It uses the subroutines PREP, READER, START, DEFAULT, GAUSS. The fortran
file DIZLIB contains several general subroutines required at various stages
in the program.
The second part is the main part and does the entire execution using 
the initial information provided by the first part. The iterative procedure
is carried out by the subroutine ITER. In the first iteration 
it solves the Dirac equation using the potential terms calculated by the
initial guessed values of the fields. The solutions (Dirac spinors) 
are used to calculate the sources which in turn  are used in solving the Klein 
Gordon equations. These solutions are used in the next iteration for the
solution of the Dirac equation. This procedure continues till the convergence
of the desired accuracy is obtained. In this operation ITER calls various
subroutines like POTGH, DIRAC, FIELD, OCCUP, DENSIT, EXPECT etc.
The output file (DIZ.OUT) is prepared in RESU and INOUT.
The user is advised to rerun the program once more after getting the
convergence. This will make the output file DIZ.OUT compact. The file DIZ.OUT
contains explicit headings to make it self-explainatory.
Several detailed comments are introduced at various places
in the program which helps the user to understand the different functions of
subroutines and at the same time to figure out what is going on at important 
steps of the program.

For an illustration DIZ.DAT file is listed for the specific case of
$^{88}$Sr. A part of the  output file DIZ.OUT is also listed providing the 
results of the RMF calculation for this nucleus, using the parameter set NL-SH.
If ones wishes to use a different set of Lagrangian parameters then the new
set of parameters can be inserted in place of the corresponding numbers of
the NL-SH set in the subroutine DEFAULT.

\bigskip

\section{\bf Acknowledgement}
Support from the Bundesministerium
f\"ur Forschung und Technologie under the project 06TM734
(6) is acknowledged. One of the authors (G.A.L) acknowledges support 
by the European Union under the contract TMR-EU/ERB FMBCICT-950216.

\newpage
\leftline{\bf Figure Captions.}
\bigskip
\bigskip
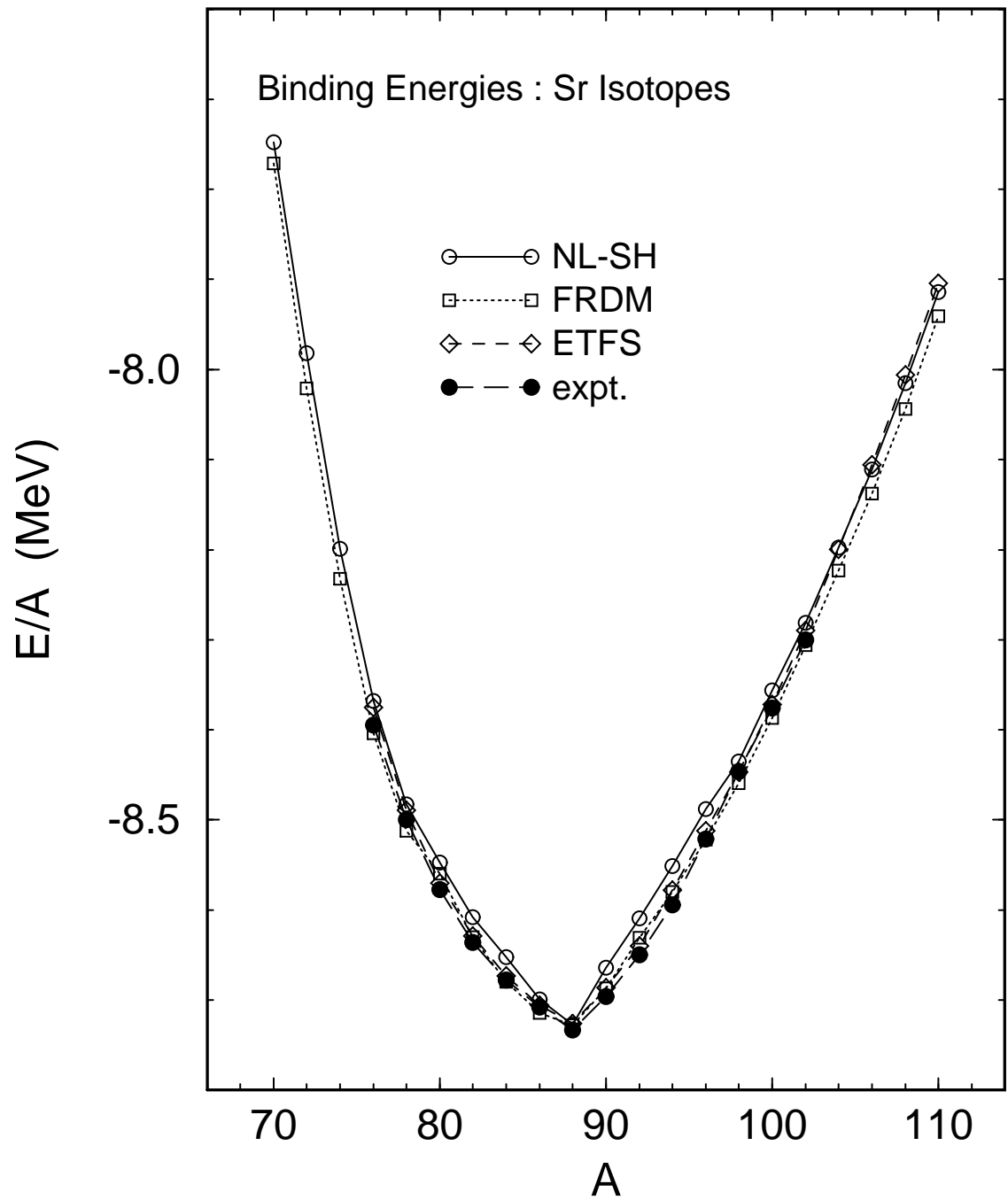
{\bf Fig. 1} The calculated RMF binding energy per nucleon for Sr isotopes 
obtained with the parameter set NL-SH. The predictions from the mass
models FRDM and ETF-SI are also shown for comparison.

\bigskip
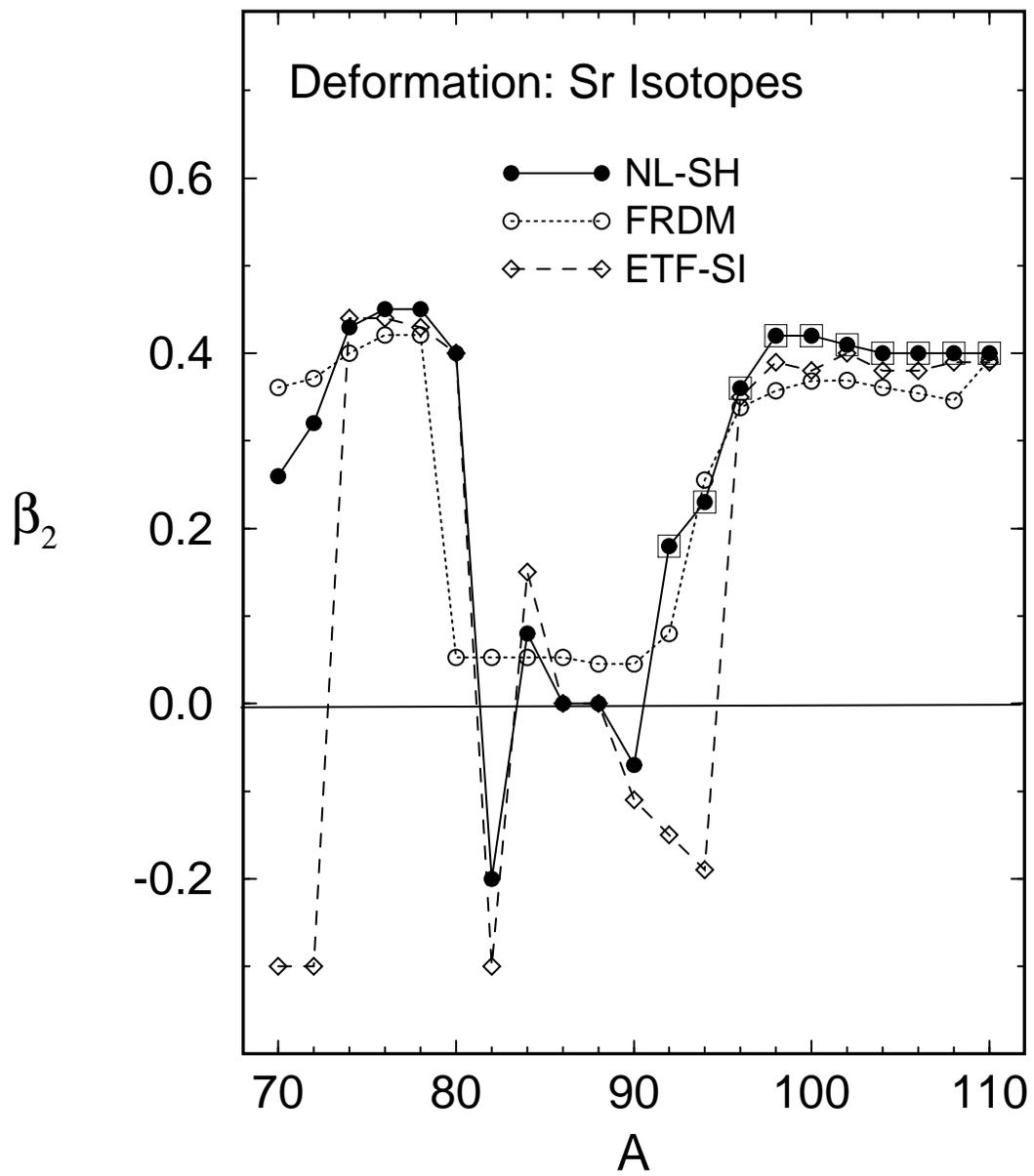
{\bf Fig. 2} The quadrupole deformation $\beta_2$ obtained from relativistic
Hartree minimization for Sr isotopes using the force NL-SH. The
predictions of the mass models FRDM and ETF-SI are also displayed for 
comparison. Nuclei exhibiting a shape coexistence and  thus a second minimum 
in the RMF theory are depicted by a square surrounding the $\beta_2$ value
of the lowest minimum. 

\bigskip
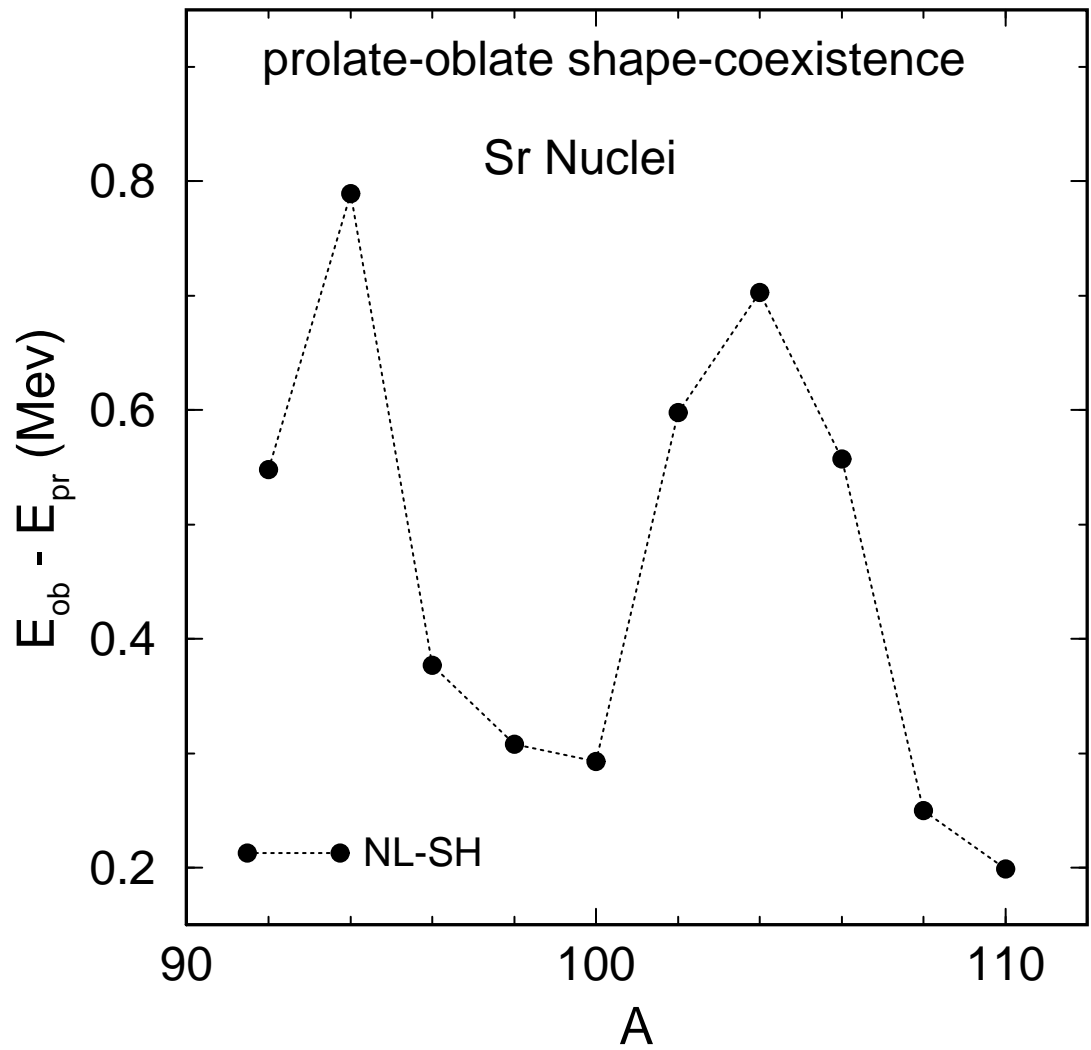
{\bf Fig. 3} The prolate-oblate shape coexistence for neutron-rich Sr
isotopes predicted in the RMF theory. The energy difference in the prolate 
and oblate minima for Sr isotopes is shown.

\bigskip
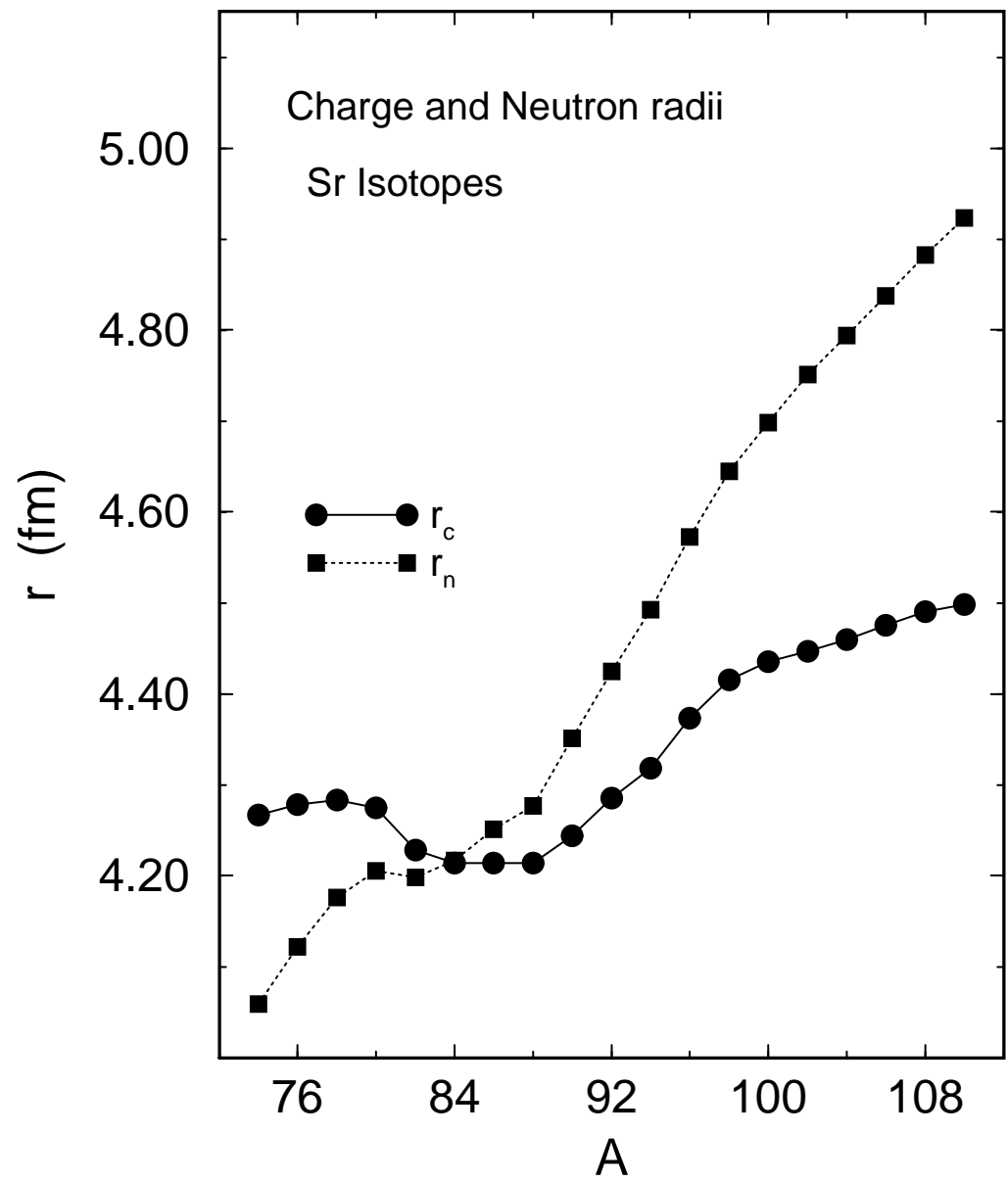
{\bf Fig. 4} The calculated rms charge and neutron radii of Sr isotopic 
chain obtained by using the force NL-SH.

\bigskip
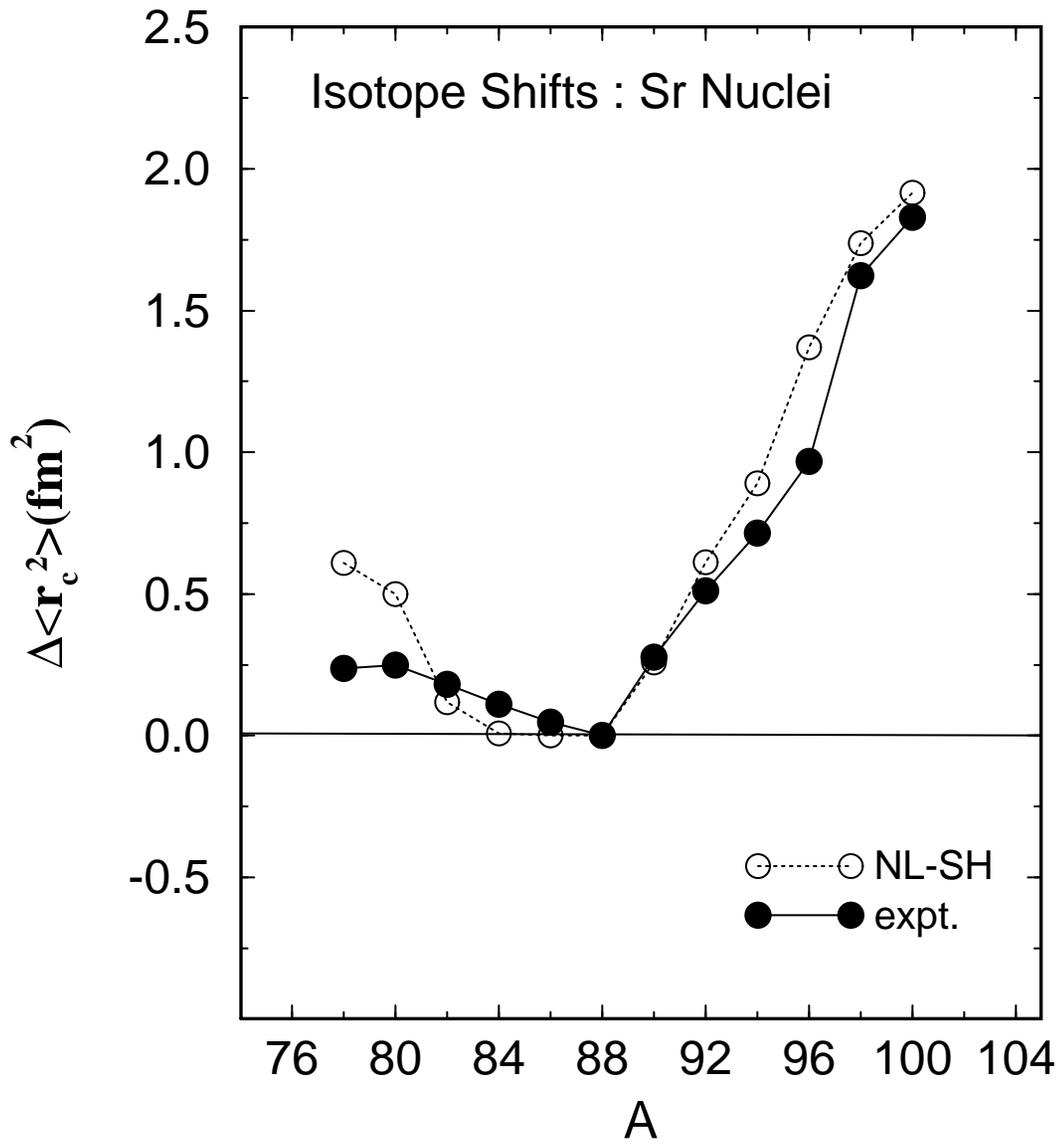
{\bf Fig. 5} The calculated and the experimental \cite{ott88,buch90} isotope 
shifts for Sr isotopes.

\bigskip
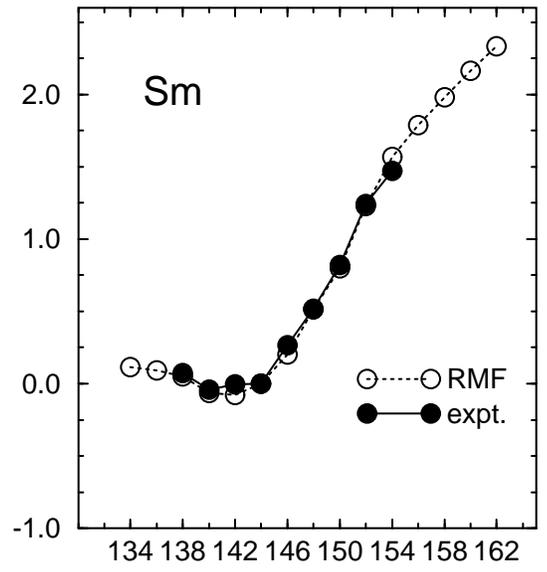
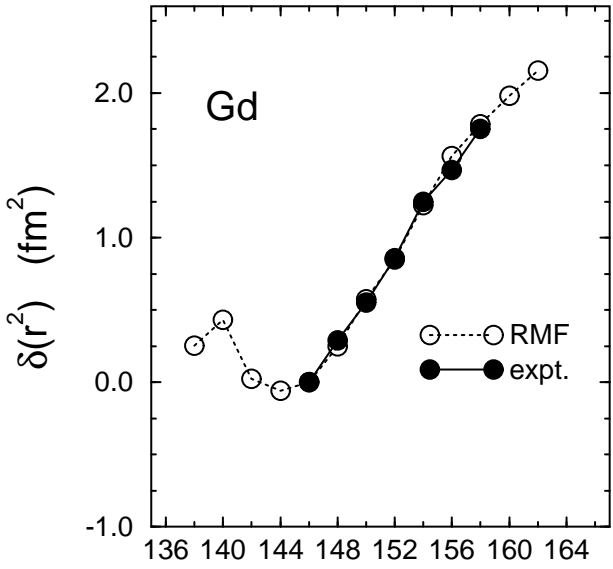
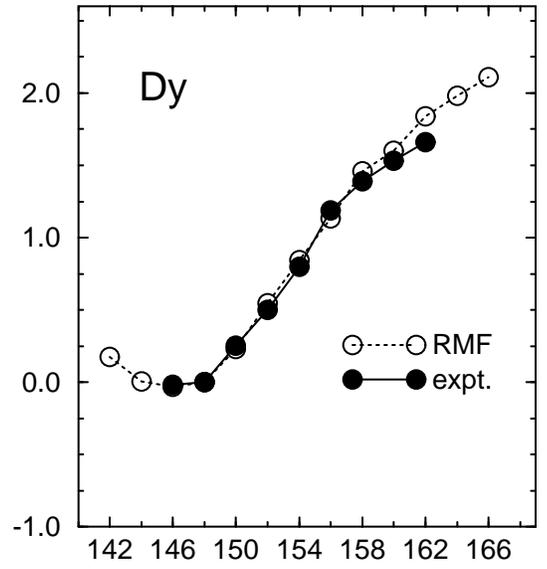
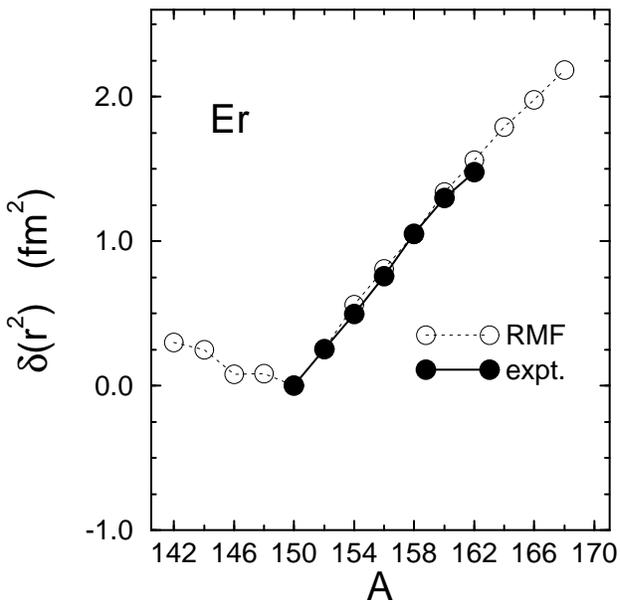
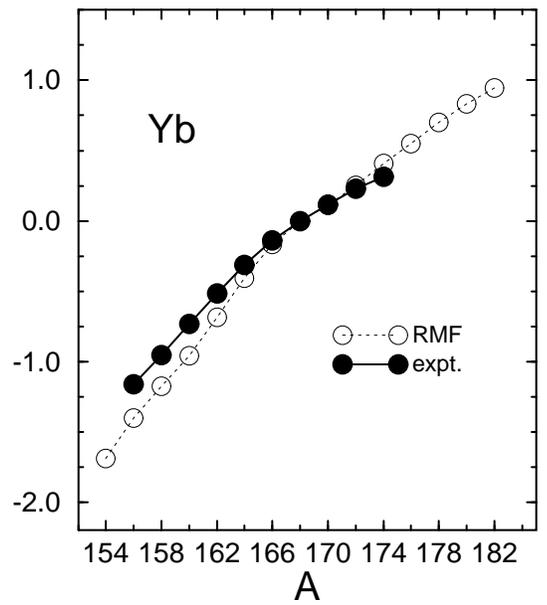
{\bf Fig. 6} The calculated and the experimental isotope shifts for Nd, Sm, 
Gd, Dy, Er, Yb nuclei. The empirical values for all but Gd nuclei are taken
from ref. \cite{ott88}. The isotope shifts for Gd nuclei have been derived
from ref. \cite{ABD.88}.

\newpage

\noindent\begin{table}
\begin{center}
\caption{\sf The parameters of the force NL-SH. All the masses are in MeV,
while $g_{2}$ is in fm$^{-1}$. The other coupling constants are
dimensionless.}
\bigskip
\begin{tabular}{ll c c c c c   l}
\hline\hline
& M = 939.0 &$m_\sigma$ = 526.059 & $m_\omega$ = 783.0 & $m_\rho$
= 763.0 &  &\\
& & & & & & \\
&$g_\sigma$ = 10.444& $g_\omega$ = 12.945& $g_\rho$ = 4.383& $g_2$ = $-$6.9099
&$g_3$ = $-$15.8337&\\ 
\hline\hline
\end{tabular}
\end{center}
\end{table}
%==========================TABLE 2============================
\noindent\begin{table}
\begin{center}
\caption{\sf The quadrupole deformations $\beta_2$ for Sr isotopes
obtained in the RMF theory using the force NL-SH. The FRDM and ETF-SI
predictions are also listed. The available empirical deformations (expt.) 
obtained from the BE(2) values are also given in the last column. 
The experimental values do not depict the signature of the deformation. 
The deformations of nuclei showing a shape coexistence with a second minimum 
are given in the parentheses.}
\bigskip
\begin{tabular}{lll c c c c c cl}
\hline\hline 
& A & NL-SH &FRDM&ETF-SI& expt.& \\
\hline
&72 & 0.324&0.371&-0.30   &  -&\\
&74& 0.430&0.400&0.44  &- &\\
&76&0.450&&0.421&0.44  & -& \\
&78&0.450&0.421& 0.43 & 0.434&\\
&80&0.402&0.053&0.40  & 0.377 &\\
&82&-0.200&0.053&-0.30 & 0.290 &\\
&84&0.089&0.053&0.15  & 0.211 &\\
&86&0.0&0.053&0.00 & 0.128&\\
&88&0.0&0.045&0.00 & 0.117&\\
&90&-0.058&0.045&-0.11 & -&\\
&92&0.181(-0.165)&0.080&-0.15 &-&\\
&94&0.230(-0.218)&0.255&-0.19 &-&\\
&96&0.356(-0.275)&0.338&0.35 &-&\\
&98&0.424(-0.309)&0.357&0.39 &0.354&\\
&100&0.426(-0.314)&0.368&0.38 &0.372&\\
&102&0.413(-0.295)&0.369&0.40 & -&\\
&104&0.403(-0.277)&0.361&0.38 & -&\\
\hline\hline
\end{tabular}
\end{center}
\end{table}  

\bigskip
\bigskip
\baselineskip = 14pt
\begin{thebibliography}{999}
\bibitem{SW.86}B.D. Serot and J.D. Walecka, 
    Adv. Nucl. Phys. {\bf 16} (1986) 1.
\bibitem{Ser.92}B.D. Serot, Rep. Prog. Phys. {\bf 55} (1992) 1855.
\bibitem{GRT.90} Y.K. Gambhir, P. Ring and A. Thimet, Ann. Phys. (N.Y.) 
{\bf 194} (1990) 132.
\bibitem{SLR.93}M.M. Sharma, G.A. Lalazissis, and P. Ring,
    Phys. Lett. {\bf B317} (1993) 9.
\bibitem{SNR.94} M.M. Sharma, M.A. Nagarajan, P. Ring, Ann. Phys. (N.Y.)
{\bf 231} (1994) 110.
\bibitem{SLH.94}M.M. Sharma, G.A. Lalazissis, W. Hillebrandt, and P. Ring; 
    Phys. Rev. Lett. {\bf 72} (1994) 1431
\bibitem{LS.95}G.A. Lalazissis and M.M. Sharma, 
    Nucl. Phys. {\bf A586} (1995) 201.
\bibitem{LSR.96} G.A. Lalazissis, M.M. Sharma and P. Ring, 
        Nucl. Phys. {\bf A597} (1996) 35.
\bibitem{PRB.87} W. Pannert, P. Ring and J. Boguta, Phys. Rev. Lett. {\bf 59}
(1987) 2420.
\bibitem{RG.88} Y.K. Gambhir and P. Ring, Phys. Lett. {\bf 202B} (1988) 5.
\bibitem{BB.77}J. Boguta and A.R. Bodmer, Nucl. Phys. {\bf A292} (1977) 413.
\bibitem{MNM.95}P. M\"oller, J.R. Nix, W.D. Myers, and W.J. Swiatecki, 
    Atomic Data and Nuclear Data Tables 59 (1995) 185.
\bibitem{AS.70} M. Abramowitz and I. A. Stegun, ``{\it Handbook of Mathematical
Functions}'', Dover, New York, 1970.  
\bibitem{GR.93} Y.K. Gambhir and P. Ring, Mod. Phys. Lett. {\bf 8} (1993) 787.
\bibitem{VAUT.73} D. Vautherin, Phys. Rev. C {\bf 7} (1973) 296.
\bibitem{DG.80} J. Decharge and D. Gogny, Phys. Rev. C {\bf 21} (1980) 1568.
\bibitem{SNR.93} M.M. Sharma, M.A. Nagarajan, and P. Ring, Phys. Lett.
                {\bf B 312} (1993) 377.
\bibitem{audi93} G. Audi and A.H. Wapstra, Nucl. Phys. {\bf A565} (1993) 1.
\bibitem{APD.95}Y. Aboussir, J.M. Pearson, A.K. Dutta, and F. Tondeur,
     Atomic Data and Nuclear Data Tables 61 (1995) 127.
\bibitem{ott88} E.W. Otten, in Nuclear Radii and Moments of Unstable Nuclei, 
               in Treaties on Heavy-Ion Science, (ed. D.A. Bromley) Vol. 7  
               (Plenum, N.Y. 1988) p. 515.
\bibitem{buch90} F. Buchinger, E.B. Ramsay, E. Arnold, W. Neu, R. Neugart,
       K. Wendt, R.E. Silverans, P. Lievens, L. Vermeeren, D. Berdichevsky, 
       R. Fleming, D.W.L. Sprung and G. Ulm, Phys. Rev. {\bf C41} (1990) 2883.
\bibitem{raman87} S. Raman, C.H. Malarkey, W.T. Milner, C.W. Nestor, and P.H. 
                  Stelson, At. Data Nucl. Data Tables {\bf 36} (1987) 1.
\bibitem{ABD.88}G.D. Alkhazov, A.E. Barzakh, V.P. Denisov, V.S. Ivanov, 
    I.Ya. Chubukov, N.B. Buyanov, V.S. Letokhov, V.I. Mishin, S.K. Sekatskii,
    and V.N. Fedoseev; JETP Lett. {\bf 48} (1988) 413.

\end{thebibliography}
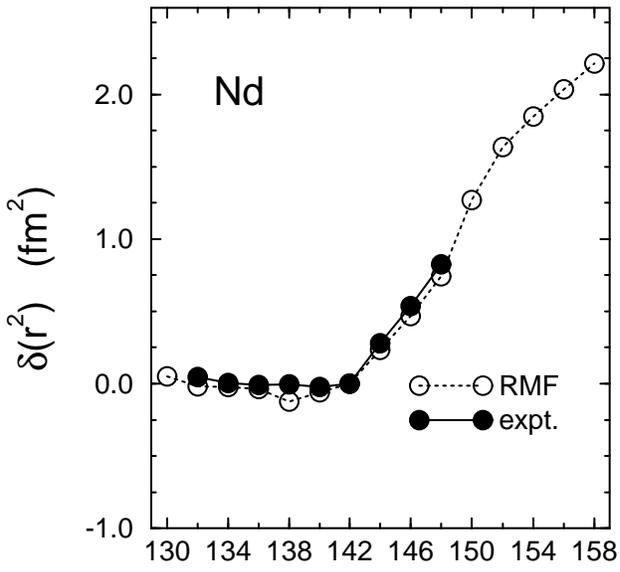
\end{document}